# Brain Controlled Wheelchair with Smart Feature

Noyon Kumar Sarkar
Electronics & Communication
Engineering Discipline
Khulna University
Khulna, Bangladesh
noyonkumarsarkar@gmail.com

Moumita Roy
Electronics & Communication
Engineering Discipline
Khulna University
Khulna, Bangladesh
moumita_m200958@ku.ac.bd

Dr. Md. Maniruzzaman
Electronics & Communication
Engineering Discipline
Khulna University
Khulna, Bangladesh
m_m_zaman@hotmail.com

*Abstract*—In Asia, many individuals with disabilities rely on wheelchairs for mobility. However, some people, such as those who are fully disabled or paralyzed, cannot use traditional wheelchairs despite having fully functioning cognitive abilities. To address this issue, we propose the development of an electric wheelchair that can be controlled using EEG signals and eye blinks. The project utilizes a MindWave Mobile device and Arduino to enable seamless control. Additionally, various sensors are incorporated to enhance the system's reliability. An ultrasonic sensor helps avoid unexpected collisions, while a smoke sensor detects hazardous smoke levels, triggering an automatic alert via a short message to a designated person. Similarly, if the passenger falls from the wheelchair, a notification will also be sent. The wheelchair's movement is controlled via an Android application, with eye-blink detection serving as the primary input method for navigation. This innovative design offers a cost-effective solution, making it accessible for widespread use. By integrating these advanced features, the system can be implemented on motorized wheelchairs to better support individuals with disabilities and enhance their independence.

## I. Introduction

The human body is controlled by the brain. The human brain contains around a hundred billion neurons. Neurons are the fundamental working unit of the brain. Motor neurons control the movement of our muscles. That's why we can move but it is very difficult for paralyzed or handicapped people to make use of motor neurons. So that they can't control their muscles and are unable to move. Neurons are also responsible for thoughts and emotions. The pattern of interaction of neurons produces electrical waves. When the thought changes the pattern of interaction of neurons also changes. So different electrical waves will produce. By sensing this electrical wave we can collect data from the brain. By using this brain wave we will make a brain-controlled prototype wheelchair [13,14]. This electrophysiological monitoring method to record electrical activity of the brain is known as Electroencephalography (EEG) [1]. There are around 32 regions of the brain lobe in our brain from where we can measure EEG signals [11].

EEG signals consist of various frequency types generated by the brain. These frequencies are categorized into five primary brain wave types, each corresponding to a distinct frequency range, collectively referred to as EEG bands. Delta has a 0 to 4 Hz frequency range which shows a deep state of sleep. Theta has a 4 to 8 Hz frequency range which shows Deep meditation & lucid dreaming. Alpha has an 8 to 12 Hz frequency range which shows Relaxation or creativity. Beta has a 12 to 32 Hz frequency range which shows Analytical thinking or Stress/Anxiety. Beta has more than 32 Hz frequency range which shows Wide brain activities or higher brain disorder.

Several wheelchairs have been designed based on the concept "brainwave controlled wheelchair". Utkarsh Sinha, Priyanka Saxena, and Kanthi M developed a wheelchair that was controlled by mind wave[2]. Debosmita Paul and Moumita Mukharjee developed a wheelchair controlled by EEG signals. The project employs a Neurosky Mindwave Mobile device to detect and transmit brainwave activity. [3,12]. Roger Achkar and his group (2015) built up an advanced mobile phone-controlled wheelchair[4]. These works motivate us to do something more with EEG signals. Android devices are available nowadays. So we thought that if we can make an android application, which will collect EEG signals from our brain and make commands for wheelchairs, it would be great for handicapped people who can't move. That's why we started to make a brain-controlled wheelchair.

Chapter 1 provides the introduction of our project. Chapter 2 describes the methodology of our project, project flow, project materials, hardware materials, software materials, hardware development, and software development. Chapter 3 gives the performance analysis of our project. Chapter 4 provides the conclusion and future research scope of this project.

## II. Methodology

In this project, we use a brain wave sensor named neurosky mindwave mobile to analyze the EEG signal. This sensor senses the RAW EEG signal. We need to decode this signal and transmit it to the car section. For this reason, we need BCI technology [5]. Brain-Computer Interface (BCI) is a technology that enables direct communication between the human or animal brain and an electronic device[6]. So we use the BCI technique to control the Arduino prototype wheelchair through brain waves.

In this project, we need a medium, which will collect the EEG signal [15]. Then process the signal and transmit it to the Arduino prototype car. That's why we created an android application. A smartphone works as our medium. When a person opens the application and connects with the brain sensor, the sensor will collect the blinking data and attention level data, then transmit it to the android application. Then the application calculates the blinking level and attention level with a threshold value. After several tests, we set the threshold value. If blinking and attention level will match with set point then the application will transmit the command to the android prototype wheelchair. Then the wheelchair will automatically move. When the blinking level crosses the set point the wheelchair will automatically stop. Figure 3 shows the whole process. In this project sometimes eye blink does not work properly. For this reason, we use an ultrasonic sensor and a buzzer. If there are obstacles in front of the prototype wheelchair and our application does not work properly then ultrasonic will measure the distance between the prototype wheelchair and obstacles. If the prototype wheelchair is so close to obstacles, the wheelchair will be stopped and the buzzer will start alarming. We use a smoke detector for the patient's safety so that if any fire or gas leak happens near the patient it will send a message via android SDK. To ensure safety during movement, a push switch is used inside the wheelchair seat. We use Bluetooth communication between brain wave sensors, smartphones, and prototype wheelchair.

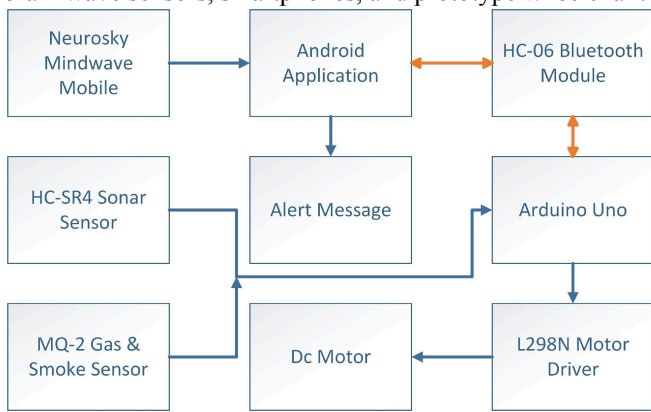

**Fig. 1**. Project System Flow.

*A. HARDWARE DEVELOPMENT*

At first, we designed our circuit diagram using Fritzing software [8]. Figure 2 shows the circuit diagram of our wheelchair. Then we defined the pin number of Arduino for each component. After that, we took a chassis and placed the component at the proper position. Then we connected them with a connecting wire.

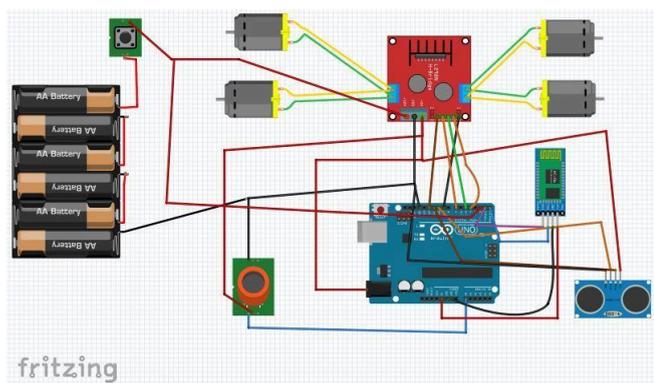

**Fig. 2.** Circuit Diagram of Prototype Wheelchair.

After making our prototype wheelchair we tested the smoke detector, ultrasonic sensor and push button. It worked properly.

*B. SOFTWARE DEVELOPMENT*

In this part, we developed an android application. It was the most important part of our project. Because this application will act as a middleman between Neurosky mindwave mobile 2 and Arduino Uno. Before generating the algorithm of the application we made a flowchart. Figure 3 shows the flow chart of the application. Android Studio is used to build our project, the Integrated Development Environment (IDE) for the Android stage [7]. It is allowed to utilize and can be downloaded over the web.

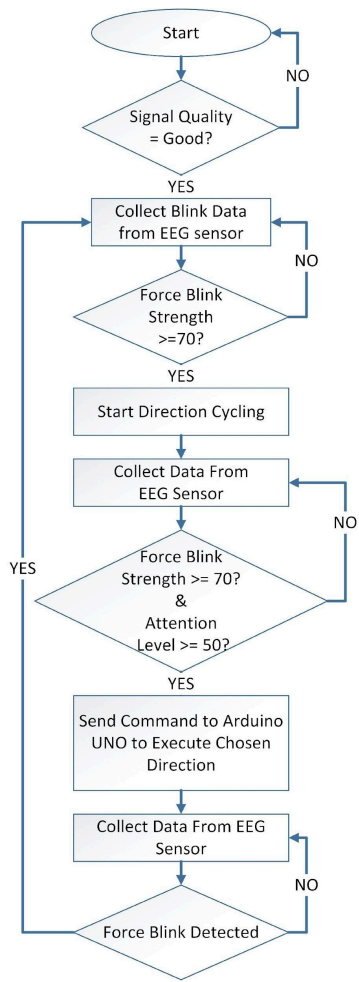

**Fig. 3.** Software System Flow.

Figure 4 shows the Android application startup GUI. The application features five buttons: one for connecting the app to the HC-06 Bluetooth module and another for connecting it to the Mindwave Mobile device. The Other 3 buttons are Fall alert, Smoke alert, and Proximity alert.

### III. RESULTS & DISCUSSION

On startup, the app recommends a mobile number in which it sends messages for those alerts as shown in figure 5. It also shows the mobile application's initial startup. The application sends the alert message using the default number which is used by the patient. Initially, the connect button for the Mindwave Mobile is disabled and becomes active only after successfully connecting to the HC-06 Bluetooth module. The app also displays six text values essential for the operation of the wheelchair.

*A. Application Test Results*

The app displays six text values that are crucial for monitoring and controlling the operation of the wheelchair. Attention level displays the user's current attention level. It needs to be above the threshold value of 50. The direction control will only be activated when the current state is set to "command." It will cycle through the forward, backward, left, and right directions, as illustrated in Figure 6. The blink strength will display values below the threshold of 70.

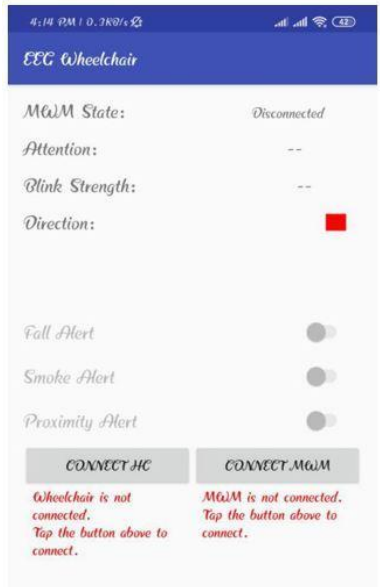

**Fig. 4.** Android application startup GUI.

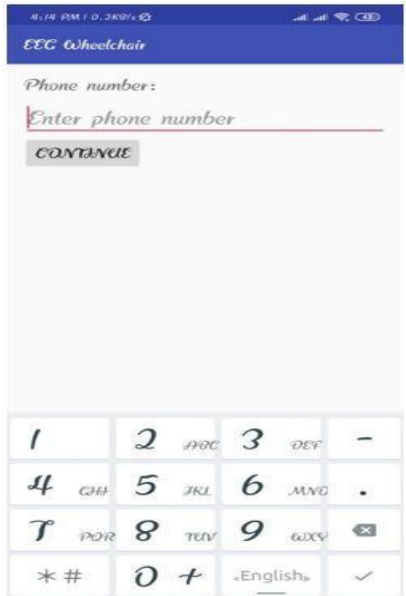

**Fig. 5**. Android Application GUI for safety purpose.

TABLE I.  CONNECTION TEST AMONG APPLICATION, MINDWAVE MOBILE AND WHEELCHAIR

| TESTE NO. | 1 | 2 | 3 | 4 | 5 |
|---|---|---|---|---|---|
| MINDWAVE MOBILE SENSOR | CONNECTED | CONNECTED | CONNECTED | CONNECTED | CONNECTED |
| HC0-6 BLUETOOTH MODULE (WHEELCHAIR) | CONNECTED | CONNECTED | CONNECTED | CONNECTED | CONNECTED |

After connecting with mindwave mobile, the application will collect our attention and blink data. Then it will compare the collected data with the defined data. If the collected data is greater or equal to define data this application will start to trigger the direction (forward, backward, right, and left) within a 2 second interval. As shown in figure 6. This is our command mode window. In this window, we will set the command by giving double blink. When a double blink event is detected, the direction cycling stops, and the current direction displayed in the cycle at the time of the double blink event becomes the selected direction. To be considered consecutive, the time interval between two blink events must be 400 milliseconds or less.

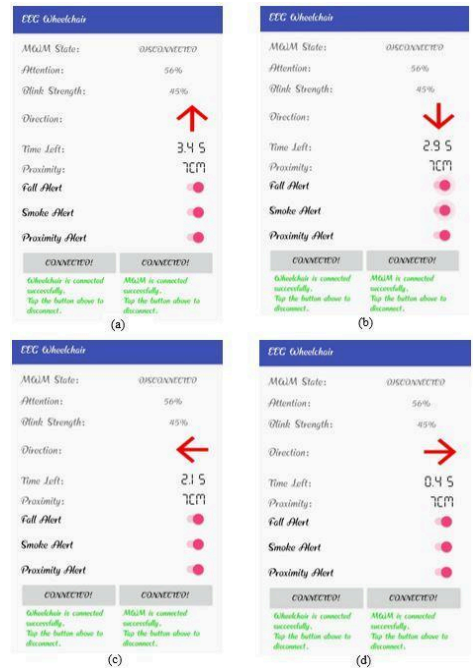

**Fig. 6.** Android application GUI direction cycle. Forward direction, Backward direction, Left direction Right direction

*B. Hardware Test Results*

We have tested all sensors and modules of our prototype wheelchair. Table II shows the results of testing of the HC-SR04 Sonar Sensor and the buzzer accuracy when an obstacle sensed in front of the wheelchair. We set the threshold value 20 cm.

TABLE II.  OBSTACLE DETECTION ACCURACY TEST

|  | Test 1 (20cm) | Test 2 (20cm) | Test 3 (20cm) |
|---|---|---|---|
| Wheelchair Status | Stopped | Stopped | Stopped |
| Buzzer Status | Buzzed | Buzzed | Buzzed |

Table III shows the results of testing of the gas sensing ability of the MQ-5 Gas Sensor and sends the message to the patient's trusted person using the android application.

TABLE III.  GAS DETECTION ACCURACY TEST

|  | Test 1 | Test 2 | Test 3 |
|---|---|---|---|
| Gas detection | Detected | Detected | Detected |
| Send message | Sent | Sent | Sent |

A patient could fall from the wheelchair and cause an accident. So to reduce accidents, we added a push button switch into the seat of the wheelchair so that if a patient falls from the wheelchair it will send a message to his/her trusted person using the android application. Table IV shows the results of testing of the push button switch.

TABLE III.  FALL DETECTION ACCURACY TEST

| | Test 1 | Test 2 | Test 3 |
|---|---|---|---|
| Fall from wheelchair detection | Detected | Detected | Detected |
| Send message | Sent | Sent | Sent |

*C. Overall Proect Data Flow*

Figure 7 shows the general data flow of this project. Data flows from the Mindwave Mobile to the Android application and, finally, to the miniature wheelchair.

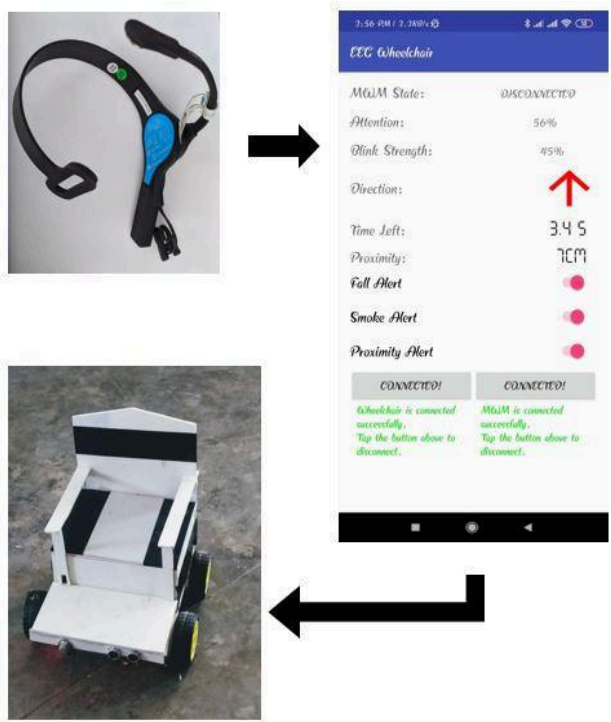

**Fig. 7.** Overall Project Data Flow.

*D. Discussion*

We completed our project work with the desired output. At the early stage of our project, we faced many problems and difficulties. Our command could not work perfectly. Motor driver IC had many problems and could not work according to our given command. It was very difficult to make the android application. Because blink data attention data was not 100% accurate. So we changed our threshold value several times while testing. Since our project didn't work properly we thought of doing something to avoid the accident. Then we connect an ultrasonic sensor and a buzzer in our project to avoid the accident. After that our prototype wheelchair stops correctly if it detects any obstacle in front of it. After trying hard we recovered all the problems and did our projects perfectly. When we gave the input command we got the expected output.

IV. CONCLUSION

This project involves recording the brain's electrical activity through a brainwave sensor, with the wheelchair's movement controlled by the user's attention level, eye blink pattern, and alpha and beta wave signals. The system integrates an engineering interface between the human brain and an Arduino-based wheelchair using an EEG brainwave sensor and firmware signal processing. The primary goal is to control the prototype wheelchair's movement in different directions—forward, backward, left, and right—using eye blinks and attention level values derived from EEG signals.

In typical operations like moving forward or turning left/right, the system can simulate various real-world control combinations, such as non-control and control states, move forward and stop, and turn while moving forward, similar to how a regular wheelchair is operated. Figure 3 illustrates the entire process.

This project enables the movement of a miniature Arduino wheelchair prototype; however, it is not flawless. While the latest EEG and brainwave technology is effective to some extent, it is far from perfect. Blink detection, in particular, is still not 100% accurate, although this issue is expected to improve as blink detection technology advances.

Due to the challenges in achieving full control over brainwaves, the inconsistencies and fluctuations in brainwave data are primarily attributed to human limitations in brainwave detection. While algorithms that process raw brainwave data can improve over time, complete and absolute control of brainwaves will remain unattainable until humans are able to better control and manipulate individual brainwave frequencies [9].

This project holds great potential for the future of EEG and brainwave-related technologies. As brainwave technology continues to improve, we can confidently predict that the day will come when EEG and brainwave-based products will be effectively integrated into everyday devices, enhancing their utility and making them more beneficial in various aspects of our daily lives.[10].

Advancements in the technology of "Mind-controlled wheelchairs" hold great promise. The performance of such wheelchairs can demonstrate their potential as an effective solution for providing independent mobility to a wide range of users who are unable to operate a traditional wheelchair system on their own. With further development, this technology can significantly enhance the quality of life for individuals with mobility impairments, offering them greater autonomy and independence.